\documentclass{svjour3}                     
\smartqed  
\usepackage{graphicx}
 \usepackage{mathptmx}      
\begin{document}

\title{One particle and Drell-Yan pair associated production\thanks{Presented at the workshop "30 years of strong interactions", Spa, Belgium, 6-8 April 2011.}}

\author{Federico Alberto Ceccopieri}

\institute{Federico Alberto Ceccopieri \at
           Dép. AGO, All\'ee du 6 ao\^ut, B\^at B5a, Universit\' de Li\`ege, 4000 Li\`ege, Belgique \\
              \email{federico.ceccopieri@hotmail.it}           
}

\date{Received: date / Accepted: date}

\maketitle

\begin{abstract}
We briefly discuss the collinear factorization formula 
for the associated production of one particle and a Drell-Yan pair 
in hadronic collisions. 
We outline possible applications of the results to three different research 
areas. 
\end{abstract}

\section{Introduction}
\label{intro}
The description of particle production in hadronic collisions 
is interesting and challenging in many aspects. Perturbation 
theory can be applied whenever a sufficiently hard scale characterizes
the scattering process. 
The comparison of early LHC charged particle spectra with 
next-to-leading order perturbative QCD predictions~\cite{stratmann}
shows that the theory offers a rather good description of data down to  
hadronic transverse momentum of the order of a few GeV. 
For hadron production at even lower transverse
momentum, the theoretical description in terms of perturbative QCD breaks 
down since both the coupling and partonic matrix elements diverge as the transverse momentum 
of final state parton vanishes.
In this paper we will study the semi-inclusive version of the Drell-Yan process, 
$ H_1 + H_2 \rightarrow H + \gamma^* + X$, in which one particle is measured in the final 
state together with the Drell-Yan pair. 
In such a process the high invariant mass of the lepton pair, $Q^2$, constitutes the perturbative
trigger which guarantees the applicability of perturbative QCD.
The detected hadron $H$ could then be used, without any phase space restriction,
as a local probe to investigate particle production mechanisms.
The evaluation~\cite{SIDYmy,SIDYmy2} of $\mathcal{O}(\alpha_s)$ corrections shows   
that there exists a class of collinear singularities 
escaping the usual renormalization procedure which amounts to reabsorbing  collinear divergences 
into a redefinition of bare parton and fragmentation functions.
Such singularities are likely to appear in every fixed order calculation
in the same kinematical limits spoiling the convergence of the perturbative 
series. In Refs~\cite{SIDYmy,SIDYmy2} a generalized procedure for the factorization of such additional collinear singularities is proposed. The latter is the same as the one proposed in Deep Inelastic Scattering~\cite{Graudenz}
where the same  collinear singularities pattern is also found, 
confirming the universality of collinear radiation between different hard processes.  
Such a generalized factorization will make use the concept of fracture functions and the corresponding renormalization group equations~\cite{Trentadue_Veneziano}. 
In a pure parton model approach, 
these non-perturbative distributions effectively describe 
the hadronization of the spectators system in hadron-induced reactions.   
The resulting transverse-momentum integrated cross-section is finite and valid for 
all transverse momentum of the detected hadron, without any 
restriction imposed by the singular behaviour of matrix elements. 
We further note that this process is the single-particle counterpart 
of electroweak-boson plus jets associated production~\cite{zjets}, 
presently calculated at nex-to-leading
order accuracy with up to three jets in the final state~\cite{w3jet}.
One virtue of jet requirement is that it indeed avoids 
the introduction of fragmentation functions to model the final state, 
which are instead one of the basic ingredients entering our formalism.
At variance with our case, however, jet reconstruction at very 
low transverse momentum starts to be challenging~\cite{cacciari}
and it makes difficult the study of this interesting portion of the produced particle spectrum.  \\
\section{Collinear facrotization formula}
\label{sec:1}
\begin{figure}[h]
\centerline{\includegraphics[width=4.5cm]{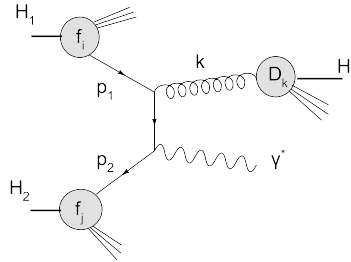}}
\caption{Example of diagram contributing to hadron production in the central
fragmentation region to order $\mathcal{O}(\alpha_s)$ in eq.(\ref{NLOcentral}).}
\label{fig1}
\end{figure}

The associated production of a particle and a Drell-Yan pair 
in term of partonic degrees of freedom starts 
at $\mathcal{O}(\alpha_s)$. One of the contributing diagrams 
is depicted in Fig.~(\ref{fig1}). 
Assuming that the hadronic cross-sections 
admit a factorization in term of long distance non-perturbative 
distributions and short distance perturbative calculable matrix elements 
for the partonic process $i(p_1)+j(p_2)\rightarrow l(k)+\gamma^*$, 
predictions based on perturbative QCD are obtained 
convoluting the relevant partonic sub-process cross-sections, $d\hat{\sigma}^{ij \rightarrow l \gamma^* }$, 
with parton distribution functions, $f_i$ and $f_j$, and fragmentation functions,
$D^{H/l}$. The hadronic cross-sections, 
at centre of mass energy squared $S$, 
can be symbolically written as~\cite{SIDYmy,SIDYmy2}
\begin{equation}
\label{NLOcentral}
\frac{d\sigma^{H,C,(1)}}{dQ^2 dz} \propto \; \sum_{i,j,l}
\int \frac{dx_1}{x_1} \int \frac{dx_2}{x_2} \int \frac{d\rho}{\rho}
 f_i^{[1]}(x_1) \, f_{j}^{[2]}(x_2)  \, D^{H/l}(z/\rho) \,
\frac{d\hat{\sigma}^{ij\rightarrow l\gamma^*}}{dQ^2 d\rho},
\end{equation}
where the convolution are over the momentum fractions  
of the incoming and outgoing partons. The variable $z$ is the energy of the observed hadron $H$
scaled down by the beam energy, $\sqrt{S}/2$, in the hadronic centre of mass
system and $\rho$ is its partonic analogue. The invariant mass of the virtual photon is indicated by $Q^2$. 
The partonic indeces $i$, $j$ and $l$ in the sum run on the available partonic sub-process. 
The superscripts label the incoming hadrons and the presence of crossed term is understood.
Within this production mechanism, the observed hadron $H$ is 
generated by the fragmentation  of the final state parton $l$, and for this reason 
we address it as \textit{central}. The amplitudes squared~\cite{DYNLO}, however,
are singular when the transverse momentum of the final state parton vanishes. 
In such configurations, the parent parton $l$ of the
observed hadron $H$ is collinear the incoming parton $i$ or $j$.
Perturbation theory looses its predictivity as these phase space 
region are approached.
The same pattern of collinear singularities are found also 
in an analogue calculation in Deep Inelastic Scattering~\cite{Graudenz}. 
In both processes such singularities can not be treated with the usual 
renormalization procedure which amounts to reabsorb  collinear divergences 
into a redefinition of bare parton and fragmentation functions. 
A widely used procedure to deal with such singularities (for example 
in modern Monte Carlo generators) is to introduce an arbitrary 
cut-off on the produced parton transverse momentum. 
It would be, however, highly desiderable 
to develop a technique to resum such logarithmic contributions to all orders in perturbation
theory since such divergences will occur in every higher order calculation
spoiling the convergence of the perturbative expansion in this kinematical limit. 
Fracture functions together with their own renormalization group equations ~\cite{Trentadue_Veneziano}
can be shown to provide the correct tool to perform such resummation. 
Bare fracture functions, $M^{H/H_1}_i(x,z)$, parametrize hadron production 
at vanishing transverse momentum. They express the conditional probability 
to find a parton $i$ entering the hard scattering while an hadron $H$ 
is produced with fractional momentum $z$ in the target fragmentation region 
of the incoming hadron. 
The use of fracture functions opens the possibility to have 
particle production already to $\mathcal{O}(\alpha_s^0)$, 
since the hadron $H$ can be non-pertubatively produced 
by a fracture function $M$ itself.
\begin{figure}[t]
\centerline{\includegraphics[width=3.5cm]{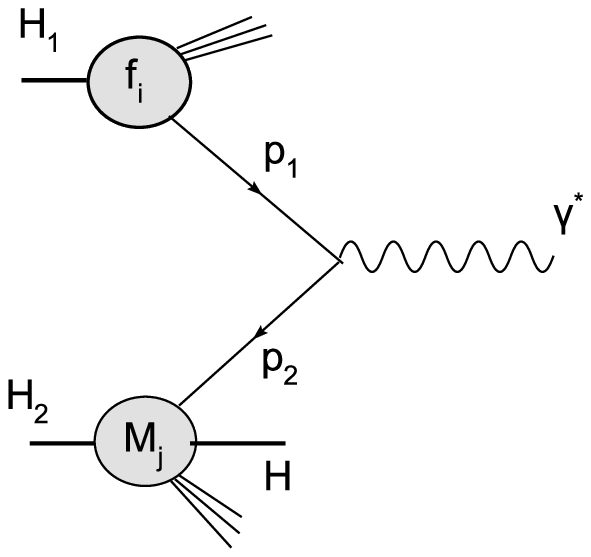}
\includegraphics[width=3.5cm]{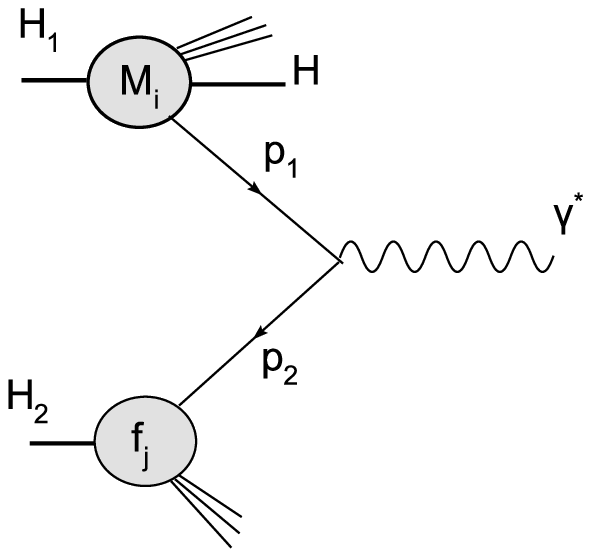}}
\caption{Parton model formula, eq.(\ref{LO}), for the associated production of a particle and a Drell-Yan pair.}
\label{fig2}
\end{figure}
The lowest order parton model formula can be symbolically written as
\begin{equation}
\label{LO}
\frac{d\sigma^{H,T,(0)}}{dQ^2 dz} \propto \sum_{i,j}
\int \frac{dx_1}{x_1} \int \frac{dx_2}{x_2} 
\big[M_i^{[1]}(x_1,z) \, f_j^{[2]}(x_2) + M_i^{[2]}(x_2,z) \, f_j^{[1]}(x_1)\big] 
\frac{d\hat{\sigma}^{ij\rightarrow \gamma^*}}{dQ^2}
\end{equation} 
where the superscripts indicate from which incoming hadron, $H_1$ or $H_2$,
the outgoing hadron $H$ is produced through a fracture functions.
This production mechanism is sketched in Fig.~(\ref{fig2}). 
So far we have only considered $\mathcal{O}(\alpha_s)$ corrections 
in the central fragmentation region, eq.~(\ref{NLOcentral}), 
to the parton model formula, eq.~(\ref{LO})\,.
In order to complete the calculation to $\mathcal{O}(\alpha_s)$ 
we should also consider higher order corrections in process initiated by 
a fracture functions. In this case, in fact, the hadron $H$ is already produced by these 
distributions and therefore final state parton in real emission diagrams should be 
integrated over and results added to virtual corrections.
One of the contributing diagrams is depicted in Fig.~(\ref{fig3}).
\begin{figure}[t]
\centerline{\includegraphics[width=3.5cm]{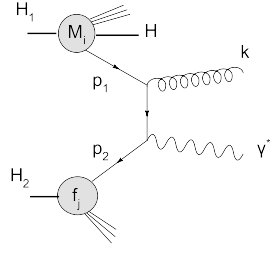}}
\caption{Example of diagram contributing to $\mathcal{O}(\alpha_s)$ corrections in the 
target fragmentation region, eq.(\ref{NLOtarget}).}
\label{fig3}
\end{figure}
The general structure of these terms is  
\begin{equation}
\label{NLOtarget}
\frac{d\sigma^{H,T,(1)}}{dQ^2 dz} \propto \sum_{i,j}
\int \frac{dx_1}{x_1} \int \frac{dx_2}{x_2} 
\big[M_i^{[1]}(x_1,z) \, f_j^{[2]}(x_2) + M_i^{[2]}(x_2,z) \, f_j^{[1]}(x_1)\big] 
\frac{d\hat{\sigma}^{ij\rightarrow l\gamma^*}}{dQ^2}\,.
\end{equation}
We refer to this corrections term as to the \textit{target} fragmentation contribution.
The calculation is, a part from different kinematics, completely  
analogue to the inclusive Drell-Yan case. 
The factorization procedure 
is accounted for by substituting in eq.~(\ref{LO}) the bare
fracture and distributions functions by their renormalized version~\cite{SIDYmy,SIDYmy2}. 
Renormalized parton distributions and fracture functions 
homogeneous terms do cancel all singularities present in eq.~(\ref{NLOtarget}).
The additional singularities in eq.~(\ref{NLOcentral}) are cancelled by the combination 
of parton distributions and fracture functions inhomogeneous renormalization terms.
Adding all the various contributions, the resulting $p_t$-integrated 
cross-sections, up to order $\mathcal{O}(\alpha_s)$, is then infrared finite~\cite{SIDYmy,SIDYmy2} 
and can be simbolically written as   
\begin{equation}
\label{NLOtot}
\frac{d\sigma^H}{dQ^2 dz} \propto \sigma_0 \sum_{i,j}
\big[M_i^{[1]} \otimes f_j^{[2]} + (1 \leftrightarrow 2)\big]
\big(1+\frac{\alpha_s}{2\pi} C^{ij}\big)
+\frac{\alpha_s}{2\pi} \; \sigma_0 \sum_{i,j,l} f_i^{[1]} \otimes f_{j}^{[2]} \otimes D^{H/l} 
\otimes K_{l}^{ij},
\end{equation}  
where $\sigma_0=4\pi\alpha_{em}^2/9 S Q^2$.
We refer to the previous equation as to the collinear factorization formula 
for the process under study.
The next-to-leading coefficients $C^{ij}$ and $K_{l}^{ij}$ have been calculated~\cite{SIDYmy2}, 
making the whole calculation ready for numerical implementation.  
In the following section we outline some applications 
of the proposed formalism. 
\section{Section Applications}
\label{sec:2}
The first possible benchmark process for the calculation 
could be strangeness production associated with a Drell-Yan pair, 
$p+p\rightarrow V+\gamma^*+X$, where $V$ generically indicates 
a $\Lambda^0$ or $\bar{\Lambda}^0$ hyperon.
The first process, according to so called leading particle effect, 
should be sensitive to the spectator system fragmentation into
$\Lambda^0$ hyperon at very low transverse momentum and therefore to be modelled with the help 
of fracture functions in $d\sigma^{H,T}$. 
In the anti-hyperon production case no such effect should be present  
so that  $\bar{\Lambda}^0$ are expected to be mainly produced by 
the fragmentation of final state parton as described by $d\sigma^{H,C}$.
The measurament of such process for different $Q^2$
could allow to test the fracture functions scale dependence embodied in their 
peculiar evolutions equations and the validity of the proposed
factorization formula in all its components. 
With this respect, a comparison of strange particle production in hadronic collisions and DIS 
would be extremely interesting since longitudinal momentum spectra of 
$\Lambda$ and $\bar{\Lambda}^0$ hyperon
has been measured quite accurately in a number of charged and neutral current DIS experiments,
both in the current as well in the target fragmentation region.

The formalism may found application in the study of  
single diffractive hard process, $p+p\rightarrow p +\gamma^*+X$, 
where the outgoing proton has almost the incoming proton energy and 
extremely low transverse momentum with respect to the collision axis. 
This process has been intensively analyzed in the DIS at HERA, 
revealing its leading twist nature. From scaling violations of the diffractive 
structure functions~\cite{H106LRG} and dijet production in the final state~\cite{H107dijet,ZEUS09final} 
quite precise diffractive parton distributions functions (DPDF) have been extracted from data, which 
parametrize the parton content of the color singlet exchanged in the $t$-channel.
The comparison of QCD predictions for single diffrative hard processes
based on diffractive parton distributions measured at HERA and assuming factorization,  
against data measured at Tevatron ~\cite{break_exp1,break_exp2},  
have indeed revealed that these processes are, not unexpectedly ~\cite{break1,break2},
 significantly suppressed in hadronic collisions, 
see the very recent analysis reported in Ref.~\cite{klasen}.
Recalling that these distributions are fracture functions in the $z\rightarrow 1$ limit, 
the present formalism can then applied to next-to-leading order accuracy, the main 
contribution to the cross-sections coming from the target term, $d\sigma^{H,T}$.
Such term can be eventually recast in triple differential form in
$x_{IP} \simeq 1-z$,  virtual photon rapidity, $y$, and invariant mass $Q^2$
and evaluated at next-to-leading order by using the appropriate coefficient 
functions~\cite{sutton}.
In this way factorization tests could be porformed at fixed $x_{IP}$ 
to avoid any Regge factorization assumption on DPDF while 
the $y$ dependence,  giving direct access to the fractional parton momentum 
in the diffractive exchange, $\beta$, 
allows to test factorization in a kinematic region which avoids 
DPDF extrapolation. Finally, the $Q^2$ dependence of the cross-sections could be used 
to investigate how factorization breaking effects eventually evolve with the hardness of the
probe and to which extent the factorized formula $M \otimes f$ actually works, as anticipated in ~\cite{break_test}.

The calculation has been performed to make predictions for 
cross-sections integrated over partonic transverse momentum.
To this end, divergent contributions, related to parton emissions 
at vaninshing transverse momentum, are factorized into fracture functions. 
In the case, however, that cross-sections are measured down to
a minimum but still perturbative hadronic transverse momentum, the latter 
constitutes a natural infrared regulator for the partonic matrix elements. 
The central production term,  $d\sigma^{H,C,(1)}$, 
can be used to estimate hadron production as the fragmentation process, 
parametrized by fragmentation functions, were happening in the QCD vacuum. 
The hadronic cross-sections can be recast in a triple differential 
form in $Q^2$, produced hadron transverse momentum $p_t$ and pseudo-rapidity $\eta$
to predict charged particle spectra or multiplicity. 
A particular interesting observable which can also be reconstructed is the 
differential cross-sections differential in $\cos \phi$, 
where $\phi$ is the angle formed by the virtual photon and the detected hadron 
in the center of mass system. This observable has been shown to be sensitive 
to the contamination of the so-called underlying event~\cite{pedestal} to jet observable 
and has been used also to investigate underlying event properties 
in Drell-Yan process~\cite{UEDY_CDF}.
Although predictions made on the present formalism take into account the radiation 
accompanying one single hard scattering per proton-proton interactions, it 
can be nevertheless used as a reference
cross-sections to gauge the impact of new phenomena, like double parton scattering.
The $d\sigma^{H,C}$ term, altough formally $\mathcal{O}(\alpha_s)$, 
is a tree level predictions and possible large higher order corrections may be 
expected especially in the forward region at large transverse momentum.
In view of this fact a $\mathcal{O}(\alpha_s^2)$ calculation at finite transverse momentum, 
as performed in DIS ~\cite{NNLO_vs_H1}, would be highly desiderable. 

\section{Conclusions}
\label{sec:3}
We have briefly reviewed a perturbative approach to single particle production 
associated with a Drell-Yan pair in hadronic collisions. 
On the theoretical side we have shown that the introduction of 
new non-perturbative distributions, fracture functions, allows a consistent factorization 
of new class of collinear singularities stemming for configurations 
in which the parent parton of the observed hadron is collinear to the incoming parton.
The scale dependence induced by this generalized factorization is driven by Altarelli-Parisi
inhomogeneous evolution equations for fracture functions which allow
the resummation to all orders of this new class of collinear logarithms.  
The factorization procedure does coincide with the one used in 
DIS confirming, as expected, the universal structure of collinear singularities among 
different hadron initiated processes and supporting the collinear factorization formula 
proposed in eq.~(\ref{NLOtot}).   
On the phenomenological side we have briefly discussed a few applications 
in which different aspects of the formalism could be tested.
The improved theoretical control on the perturbative component indeed allows the investigation 
on new phenomena appearing in hadronic collisions, for example the rapidity gap probability 
suppression in hard diffractive processes respect to diffractive DIS and 
the investigation, although indirect, of multiple parton-parton interactions.

\end{document}